\documentclass[aps,twocolumn,showpacs]{revtex4}

\usepackage{graphicx}
\usepackage{dcolumn}
\usepackage{bm}

\begin{document}
\title{Unidirectional quantum walk of two correlated particles: 
Separating bound-pair and unbound wavepacket components}
\author{A. R. C. Buarque$^1$ and W.~S.~Dias$^1$}

\affiliation{Instituto de F\'{\i}sica, Universidade Federal de Alagoas,
57072-970 Macei\'o, AL, Brazil}

\begin{abstract}
We study the unidirectional transport of two-particle quantum wavepackets in a 
regular one-dimensional lattice. We show that the bound-pair state component 
behaves differently from unbound states when subjected to an external pulsed 
electric field. Thus, strongly entangled particles exhibit a quite distinct 
dynamics when compared to a single particle system. With respect to centroid 
motion, our numerical results are corroborated with an analytical expression 
obtained using a semi-classical approach. The wavefunction profile reveals that 
the particle-particle interaction induces the splitting of the initial 
wavepacket into two branches that propagate with specific directions and drift  
velocities. With a proper external field tunning, the wavepacket components can 
perform an unidirectional transport on the same or opposite directions. The 
amplitude of each mode is related to the degree of entanglement betweem 
particles, which presents a non-monotonic dependence on the interaction 
strength.

\end{abstract}

\maketitle

\section{Introduction}

Unlike a classic random walker,  a quantum walk can be in a coherent 
superposition of several positions and exploring multiple trajectories over an 
{\it n}-dimensional graph. In periodic systems, for example, the quantum 
particle propagates much faster (ballistic propagation) than its classical 
counterpart (diffusive propagation)~\cite{kempe}. Quantum walkers have been
widely studied in a variety of different settings such as in the 
development of quantum algorithms~\cite{Shenvi,Chakraborty,Portugal}, efficient 
energy transfer in proteins complex~\cite{Mohseni}, classical 
optics~\cite{Schreiber,Grafe}, waveguide lattices~\cite{Perets,Peruzzo}, 
nuclear 
magnetic resonance~\cite{Du}, quantum dots~\cite{Manouchehri}, trapped atoms in 
optical lattices~\cite{Karski,Preiss}, disorder~\cite{Yin,Jackson}, interacting 
particles~\cite{Wang,Peixoto} and bacteria behavior in biological 
systems~\cite{loyd}.

Despite the recent progress, many investigations on quantum walks 
are related to a single walker, primarily in the experimental scope. On the 
other hand, quantum effects are considerably enhanced in systems with more than 
one walker. 
In such cases, generalizations that consider many interacting 
walkers can provide useful information regarding to universal and 
efficient quantum computation~\cite{Childs}. 
Interaction between walkers typically results in the appearance of entanglement 
and have been shown to improve certain aspects, such as in the graph 
isomorphism problem~\cite{Berry}. Quantum walks of interacting and 
non-interacting quantum particles are fundamentally different in the context of 
solving the computational problem of graph isomorphism. In this case, it was 
reported that two interacting bosons are more powerful than single particles 
and two non-interacting particles for distinguishing non-isomorphic strongly 
regular graphs~\cite{Gamble}. Quantum walks of two identical photons 
revealed quantum correlations that depended critically on the input state of 
the quantum walk~\cite{Peruzzo}. Fundamental effects such as the emergence of 
correlations in two-particle quantum walks were recently reported for 
interacting atoms in an optical lattice~\cite{Preiss}. The control over the 
interacting atoms in the regime where the dynamics is dominated by 
interparticle 
interactions made it possible to observe the frequency doubling of Bloch 
oscillations, predicted for electron systems~\cite{Dias} and recently simulated 
with photons in a waveguide array~\cite{Corrielli}.

Within the studies involving quantum dynamics under the influence of external 
fields, the unidirectional transport  of wavepackets promoted by superposed 
static and harmonic fields has been 
greatly explored~\cite{Thommen,Ivanov,Haller,Caetano,DiasACDC,Zheng}. A 
weakly interacting Bose-Einstein condensate of $Cs$ atoms in a tilted lattice 
potential was used to demonstrate that harmonic driving can lead to 
matter-wave transport over macroscopic distances~\cite{Haller}. This directed 
motion is promoted when the frequency of the AC field is a multiple of the 
frequency of the usual Bloch  oscillations. Furthermote, the average velocity 
displayed by the wavepacket depends on the magnitudes of the AC and DC field 
components and on the initial phase of the AC field. Super Bloch oscillations 
and breathing take place with an amplitude that diverges as the resonance 
condition is approached~\cite{Caetano}. For initially localized and 
uncorrelated 
two-particle quantum wavepackets evolving in a 1D discrete lattice, it has 
been 
reported that the particles become strongly entangled when directed by a 
harmonic AC field which is resonant with frequency-doubled Bloch oscillations 
promoted by a static DC field~\cite{DiasACDC}. These theoretical and 
experimental works have shown the possibility of using
external fields to manipulate entangled matter-states.

The aim of this paper is to propose a new protocol for manipulating quantum 
matter-states of two correlated particles in a regular 
one-dimensional (1D) lattice. More specifically, we consider two interacting 
particles placed in a linear discrete lattice under the effect of a Gaussian 
time-dependent electric field. Pulsed external fields (including 
Gaussian-Pulses) have been used in different scientific 
contexts~\cite{Urano,Moore,Hu,Dong,Beebe,Arlinghaus}, making the experimental 
implementation feasible. In the framework of the tight-binding 
Hubbard 
model Hamiltonian with on-site inter-particle interaction, we will
follow the time evolution of wavepackets and compute typical physical 
quantities  
to 
characterize the wavepacket dynamics along the chain. In particular, we will 
show that a
proper tunning of 
the pulsed electric field can control the migration of a pair of strongly 
entangled particles. 
Our numerical results are corroborated with an analytical expression obtained 
using a semi-classical approach. Since the external field acts in a different 
way on bound-pair and unbound states, we will show that the initial wavepacket splits into 
two branches which propagates in specific directions and drift velocities. 
The amplitude of each mode will be related to the degree of entanglement of the 
two particles, which exhibits a non-monotonic dependence on the interaction 
between particles. As such, we explore the possibility of controlling the direction 
and drift velocity of two distinct  
wavepacket components.

%

\section{MODEL AND FORMALISM}

The system under consideration contains two interacting 
particles placed in a one-dimensional discrete 
lattice of spacing $a$ under the action of an external field. In 
the framework of the tight-binding Hubbard model, the Hamiltonian can be 
described as
\begin{eqnarray}
H&=&\sum_n\sum_{s=1,2}J(c^{\dagger}_{n+1,s}c_{n,s}+c^{\dagger}_{n,s}c_{n+1,s}
)\nonumber \\
&+&\sum_n\sum_{s=1,2}\left[\epsilon_nc^{\dagger}_{n,s}c_{n,s} - eF(t)na\right] 
\nonumber \\
&+& 
\sum_n 
Uc^{\dagger}_{n,1}c_{n,1}c^{\dagger}_{n,2}c_{n,2}  ,
 \label{eq:Hamiltonian}
\end{eqnarray}
where $c_{n,s}$ and $c^{\dagger}_{n,s}$ are the annihilation and creation 
operators for particles of charge $e$ at site $n$ with spin $s$, $J$ is the 
nearest-neighbor hopping amplitude, $\epsilon_{n}$ is the potential at site 
$n$ and $U$ is the on-site Hubbard interaction. A possible physical realization 
of the present model consists of two bosonic atoms in an optical lattice under a 
tilting pulse. It has been recently shown experimentaly that two interacting 
bosons in a tilted optical lattice~\cite{Preiss} has similar features of two 
interacting charged particles under an external uniforme field~\cite{Dias}. 
Here, the 
external field consists of a Gaussian-pulse applied parallel 
to the chain lenght, which can be expressed as
\begin{equation}
 F(t)=B(\rho)exp\left[ -\frac{(t-\tau)^2}{4\rho^2}\right],
 \label{pulse}
\end{equation}
where $\rho$ controls the duration of each pulse and $\tau$ is the time of 
reference. For very small values of $\rho$, $F(t)$ represents a delta-like pulse 
at $t\approx \tau$. Pulsed 
external fields (including 
Gaussian-Pulses) have been reported in different settings such as reorientation 
of nematic liquid crystals~\cite{Urano}, the experimental realization of a 
quantum $\delta$-kicked rotor in ultracold sodium atoms trapped  in a 1D 
potential~\cite{Moore}, analysis of the transient membrane response for 
cells~\cite{Hu}, deceleration and bunching of cold molecules~\cite{Dong} and 
wave-packet 
manipulation using pulses with a smooth envelope~\cite{Arlinghaus}.

In the following, we will consider the case $U>0$, corresponding to 
Hubbard repulsion. In order to follow the time evolution of wavepackets, we 
solved the time 
dependent Schr\"odinger equation by expanding the wavefunction in the Wannier 
representation 
$|\Phi(t)\rangle=\sum_{n_1,n_2}f_{n_1,n_2}(t)|n_1,1;n_2,2\rangle$ where the 
ket $|n_1,1;n_2,2\rangle$ represents a state with one particle with spin 
$1$ at site $n_1$ and the other particle with spin $2$ at site $n_2$. We 
consider the particles distinguishable by their
spin state. Once 
the initial state is prepared as a direct product of states, the particles will 
always be distinguishable by their spins since the Hamiltonian does not involve 
spin exchange interactions. The temporal evolution of the wavefunction 
components in the Wannier representation is governed by the time-dependent 
Schr\"odinger equation
\begin{eqnarray}
i\frac{df_{n_1,n_2}(t)}{dt}&=&f_{n_{1}+1,n_{2}}(t)+f_{n_{1}-1,n_2}(t)\nonumber 
\\ &+& f_{n_{1},
n_{2}+1}(t)+ f_{n_{1},n_{2}-1}(t)\nonumber 
\\
&\hspace*{-3.7cm}+&\hspace*{-2.0cm}\left[\epsilon_{n_{1}}+\epsilon_{n_{2}}
-F(t)(n_1+n_2)+\delta_{ n_1 , n_2} U\right ]f_{n_{1},n_{2}}(t),
\label{eq:EStime}
\end{eqnarray}
where the on-site energies $\epsilon_n$ were taken as the reference energy 
($\epsilon_n=0$) and we used units of $\hbar=J=e=a=1$.  The above set of 
equations were solved numerically by using fourth order Runge-Kutta method with 
step size about $10^{-4}$ in order to keep the wavefunction norm conservation 
along the entire time interval considered. We followed the time evolution of an 
initially Gaussian wavepacket with width $\sigma$: \begin{eqnarray}
 \langle 
n_1,1;n_2,2|\Phi(t=0)\rangle=\frac{1}{A(\sigma)}e^{-\frac{
(n_1-n_1^0)^2}{4\sigma^2}}e^{-\frac{(n_2-n_2^0)^2}{4\sigma^2}},
\end{eqnarray}
where the initial positions ($n_1^0,n_2^0$) were considered to be centered at 
($N/2-d_0, N/2+d_0$). Through the above-described approach, we computed typical 
quantities that can bring information about the wavepacket 
time-evolution, as it will be detailed below.

\section{RESULTS AND DISCUSSION} 

\begin{figure}[!t]
\begin{center}
\resizebox{10cm}{10cm}{\includegraphics{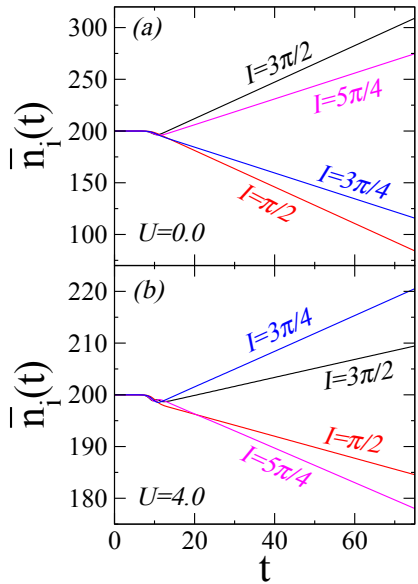}}
\caption{Time evolution of the average position of particle 1 for four settings 
of electric field pulses applied at $\tau=10$ time units. 
The resulting impulses of the electric pulses are $I=3\pi/2, 5\pi/4, 3\pi/4, 
\pi/2$. The dynamic behavior displayed by particles with interaction 
strength $U=4.0$ is clearly distinct from that provided by non-interacing 
particles ($U=0$).	
}
\label{fig1}
\end{center}
\end{figure}

We start following the time 
evolution of the wavepacket centroid associated with each particle defined as
\begin{equation}
 \bar{\tt n}_i(t)=\sum_{n_1,n_2} (n_i)|f_{n_1,n_2}(t)|^2,\hspace{.2cm}i=1,2.
\end{equation}
Due to the symmetry of the initial state and interaction Hamiltonian, one has 
that $\bar{\tt n}_1(t)=\bar{\tt n}_2(t)$. In fig.~\ref{fig1} we plot the 
centroid evolution for an initial wave-packet width $\sigma=1.0$, $d0=0$ and 
(a) $U=0.0$ and (b) $U=4.0$. We adjusted
the value of $B(\rho=1)$ in order to 
apply an electric pulse at time $\tau=10$ whose impulse 
$\left[ I=\int_{-\infty}^{\infty} {\bf F}(t)dt\right]$ was  $I=3\pi/2; 
5\pi/4; 3\pi/4$ or $\pi/2$. We observe that a single pulse with impulse $I$ 
promotes the movement of the wave-packet. However, the dynamic behavior 
displayed by particles with interaction strength $U=4.0$ is clearly distinct 
from that provided by non-interacing particles ($U=0$). While $U=0.0$ shows 
trends consistent with the single-particle formalism 
(see ref.~\cite{Silva}), for $U=4.0$ some electric pulse settings imposes a 
movement in the opposite direction to that observed for $U=0.0$. 

\begin{figure}[!t]
\begin{center}
\resizebox{10cm}{10cm}{\includegraphics{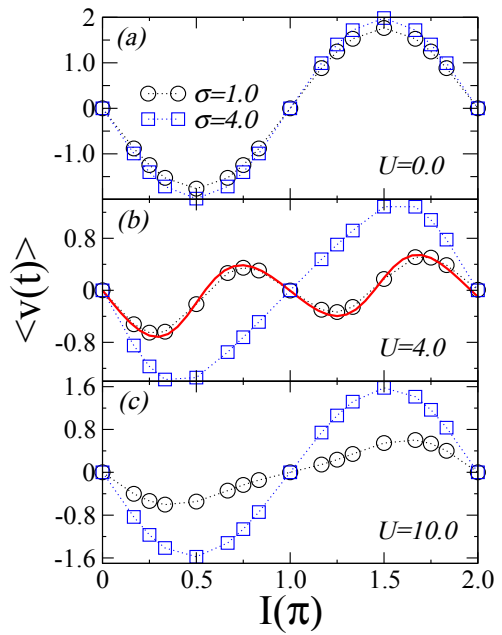}}
\caption{Impulse dependence of the centroid velocity 
for (a) $U=0.0$, (b) $U=4.0$ and (c) $U=10.0$ and two distinct initial wave 
packets $\sigma=1.0,4.0$.  For $U=0.0$ the sine-like
behavior is in good agreement with the semi-classical prediction for 
non-interacing particles. On the other hand, for an intermadiate interaction 
strength, the bound 
states seems to play a predominant role on the dynamics of particles. The solid 
line corresponds to the semi-classical dependence for 
particles performing coherent hoppings.}
\label{fig2}
\end{center}
\end{figure}

\begin{figure*}[!ht]
\begin{center}
\resizebox{15cm}{9.5cm}{\includegraphics{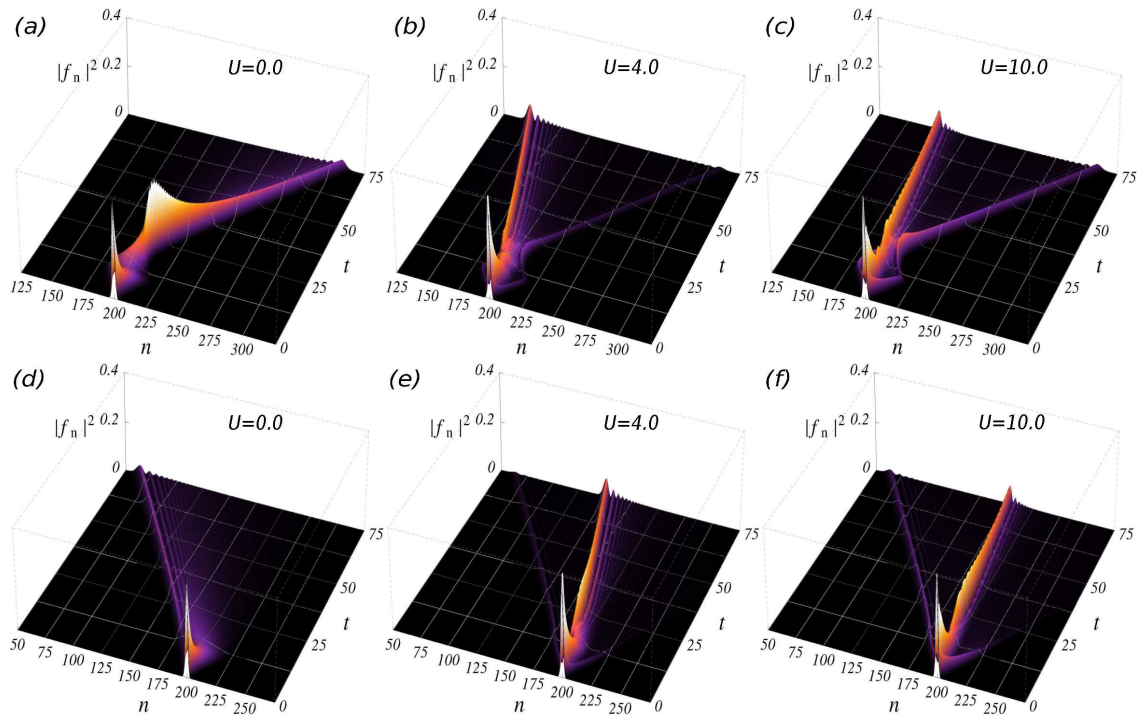}}
\caption{Time evolution of the one-particle wavefunction profile in the absence 
of interaction ($U=0.0$), intermediate 
($U=4.0$) and strong particle-particle interaction ($U=10.0$). Electric 
pulse applied at $\tau=10$ time units whose resulting impulse are: top panels 
$I=5\pi/4$ and bottom panels $I=\pi/3$.
While $U=0.0$ results in a behavior consistent with the single-particle 
formalism, the interaction induces a competition of bound-pair and unbound 
states, which associated with the electric pulse, splits the wavepacket in 
two parts. Both components (bound-pair and unbound states) perform 
an unidirectional transport, but may propagete along (a-c) opposite or (d-f) the 
same direction.}
\label{fig3}
\end{center}
\end{figure*}

In order to  better characterize the wavepacket dynamics, we collected the 
centroid after 
different electric pulse settings. With these data, we 
computed the average centroid velocity $\langle v(t)\rangle$ for each value of 
the field 
impulse. We plot in fig.~\ref{fig2} $\langle v(t)\rangle$ versus $I$ (in $\pi$ 
units) for initial wavepackets width $\sigma=1.0; 4.0$ and (a) $U=0.0$, (b) 
$U=4.0$ and (c) $U=10.0$. As the previous results showed, the behavior for 
$U=0.0$ (see 
fig.~\ref{fig2}a) recovers the single particle 
dynamics~\cite{Silva}. A clear understanding of the underlying physical process 
can be reached by using a semi-classical 
formalism. For a particle of charge $e$, the wavevector $k$ after an 
electric pulse is
\begin{equation}
k=k_0+\frac{e}{\hbar}\int_{t_i}^{t_f}F(t)dt.
\end{equation}
Thus, keeping in mind the energy dispersion of the single particle problem and 
its relationship with the group velocity of wave-packet centered around some 
$k'$ state, the sine-like behavior displayed in fig.~\ref{fig2}a is easily 
recovered. 
By increasing of the initial wavepacket  width, the dynamics converges to 
semiclassical prediction, with limits $\pm 2Ja/\hbar$. For a narrow initial 
wavepacket ($\sigma=1.0$) 
and intermediate interaction strength $U=4.0$ (see fig.~\ref{fig2}b), $\langle 
v(t)\rangle$ displays a distinct dependence on the field impulse. Now, 
besides the pure (unbound) states covering the range $-4J\leq E \leq 4J$, there 
is a band of bound-pair states covering the range $U\leq E\leq 
\sqrt{U^2+16J^2}$~\cite{Dias,claro}. These bound-pair states play a predominant 
role in 
the wavepacket dynamics~\cite{Peixoto,Preiss,Dias,claro,Shepelyansky,Dias2}. By 
assuming a transformation to the center-of-mass coordinate 
$f(n_1,n_2)=e^{ik(n_1+n_2)a}\chi(n_1-n_2)$ it is possible to show (more 
technical details are found in ref.~\cite{claro}) that in the absence of 
interaction $U$ we have 
\begin{eqnarray}
E=4Jcos(ka)cos(za).
\label{eq:enefreestates}
\end{eqnarray}
Here, $k$ is the center of mass momentum and $z$ is the relative momentum 
between particles. Besides, we have in the presence of Hubbard interaction
\begin{eqnarray}
 E=\sqrt{U^2+16J^2cos^2(ka)},
 \label{eq:boundstates}
\end{eqnarray}
related to the bound-pair states. With 
these last two expressions, since the group velocity of a wavepacket is 
$v(k)=\frac{1}{\hbar}\partial E(k)/\partial k$, we obtain
\begin{equation}
v(k)=\gamma\frac{ sin(ka)cos(ka)}{
\sqrt{U^2+16cos^2(ka)}}-\beta sin(ka)cos(za),
\end{equation}
where $\gamma$ and $\beta$ are constants.  We fitted the data in 
fig.~\ref{fig2}b with the above expression 
 (represented by the solid line) and achieved an excellent agreement of the 
numerical data with our semi-classical prediction for strongly 
correlated particles. With increasing of the initial wavepacket width, the 
unbound states components predominates, which considerably reduces 
the correlation between particles. While the interaction favors the 
coherent hopping associated with bound-pair states,  the double occupancy 
probability decreases for wide wavepackets. 
This characteristics becomes clear when we observe that  the impulse dependence 
of the average 
velocity gets closer of the non-interacting particles case for strong Hubbard 
interaction (see fig.~\ref{fig2}c).



In order to exemplify the competitive character presented above it is shown 
in fig.~\ref{fig3} the time evolution of the one-particle wavefunction 
profile in the absence of interaction ($U=0.0$), intermediate 
($U=4.0$) and strong particle-particle interaction ($U=10.0$).  The electric 
pulse was applied at $\tau=10$ time units, whose resulting impulse are: top 
panels 
$I=5\pi/4$ and bottom panels $I=\pi/3$. For both electric field 
configurations, $U=0.0$ shows that the entire wavepacket is driven to 
perform a unidirectional transport along the chain, consistent with the 
single-particle formalism. On the other hand, the electric 
pulse on the system with the interaction between particles turned on induces the 
splitting of 
the wavepacket in two wavefronts. One is associated to unbound states, 
while the other is governed by bound-pair states. 
A similar behavior was reported for correlated bosons in optical lattices 
in the presence of doubly modulated AC-fields~\cite{Zheng}. The on-site 
interaction and the lattice shaking displays independent modulating 
frequencies, that when properly adjusted can offer a bifurcating quantum motion 
of 
 pair correlated particles propagating in opposite directions. However, 
we show that both components (bound-pair and unbound components) 
performs an unidirectional transport, but can propagate in the opposite [see 
fig.~\ref{fig3}(a-c)] 
as well as the same [see fig.~\ref{fig3}(d-f)] direction. 
The drift velocity of each front can be analytically determined by the 
semi-classical description of the impulse 
dependence of the centroid velocity. 
\begin{figure}[!t]
\begin{center}
\resizebox{6cm}{9cm}{\includegraphics{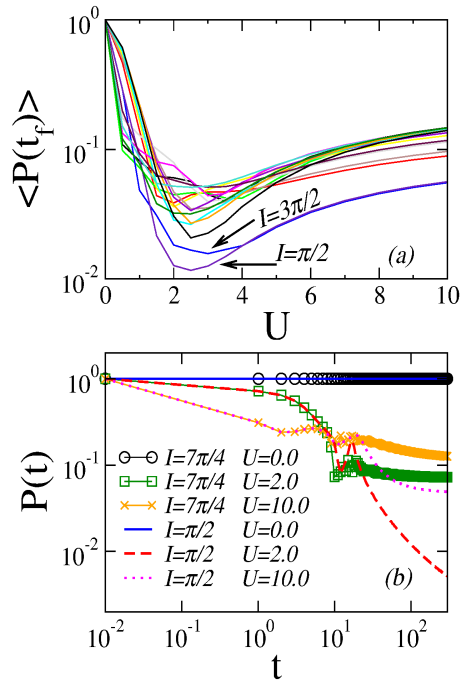}}
\caption{(a) Purity function computed after $100$ time units 
($\langle P_{12}(t_f)\rangle$) 
versus interaction ($U$) for distinct applied electric pulses. The degree of 
entanglement displays a non monotonic dependence on $U$ and is greater for 
electric pulses which promote higher velocities of the unbound branch. (b) 
For an intermediary interaction strength the time evolution of quantum purity 
function indicates that the wavepacket develops a continuously increasing 
degree of quantum entanglement when the electric pulse promote higher velocity 
of the unbound branch.}
\label{fig4}
\end{center}
\end{figure}

The previous results suggest that the connection between interaction 
strenght and the amplitude of the wavepacket fractions is related to 
competitive character between bound-pair and unbound states. 
This feature indicates that states are more strongly 
entangled in the regime of intermediate interaction strengths. In order to quantify 
the degree of entanglement of the two-particle wavefunction, we compute the purity 
function defined as
\begin{equation}
 P(t)=tr\rho_1^2(t),
\end{equation}
where $\rho_1(t)$ is the reduced density matrix for particle 1 obtained after 
taking the partial trace over the states of particle 2~\cite{Islam}. In 
fig.~\ref{fig4}a it is shown the purity computed after $100$ time units 
($\langle P_{12}(t_f)\rangle $) versus the interaction strenght ($U$) for 
systems under different applied field pulses. For a pure quantum state, the 
density matrix is a projector, so that the purity function $P_{12}(t)=1$ and the 
two particles are not quantum entangled. As the interaction between particles is 
increased, an enhancement in the degree of entanglement $(P_{12}(t_f)\rightarrow 
0)$ is observed. However, this behavior is not monotonic. After an intermediate
interaction strength, the degree of entanglement is reduced. This non-monotonic 
behavior is related to the competitive character described above, where the 
interaction favors the coherent hopping associated with bound states while the wavepacket  
widening decreases the double occupancy probability. We also note that the 
degree of entanglement is larger for electric pulses that promote larger 
velocities at the unbound branch. Fig.~\ref{fig4}b displays the time 
evolution of the purity function for $U=0, 2, 10$ and electric field impulses 
$I=7\pi/4, \pi/2$. While the two particles are not entangled for $U=0$, 
for $U=2$ and  $I=\pi/2$ the  purity function continuously decreases in 
time. This feature indicates that the wavepacket develops quantum entanglement 
over a continuously growing chain segment. In this configuration the electric 
pulse promotes the largest velocity of the unbound branch. In contrast, for other 
settings, the degree of entanglement quickly saturates, 
reinforcing the fact that unbounded states plays a more significative role in the wavepacket 
dynamics in the regime of very strong interactions.

\section{SUMMARY AND CONCLUSIONS}

In summary, we introduced a scheme to generate and manipulate spatially entangled 
two-particle states by driving them using a pulsed electric field. More 
specifically, our results showed that bound-pair and unbound states of an 
initially entanglement wavepacket can be controlled separately. This allows 
to split the wavepacket into two fractions that develops unidirectional transport, 
with speed and direction of each branch externally controlled by the 
pulsed field. The electric field can be adjusted in order to make these two
components (bound-pair and unbound states) propagate either in the same or 
opposite directions. 
The amplitude of each mode is related to the degree of 
entanglement of the two particles, which presents a non-monotonic dependence 
with the interaction between particles. This behavior comes from the 
competitive character between bound-pair and unbound states. 
This competition leads to an 
optimal range of couplings to obtain a strongly correlated dynamics of the two 
interacting particles. Our analysis is based on observables that can be 
experimentally verified, such as the wavepacket profile~\cite{Preiss} and quantum
purity~\cite{Islam}. Thus, these recent experimental achievements strongly indicate
that the scheme proposed here is feasible in systems of ultracold 
atoms trapped on one-dimensional optical lattices under a tilting pulse. We hope that our 
work may impel further 
investigations aiming the manipulation of entangled particles in low-dimensional 
systems.

\section{Acknowledgments}

We would like to thank M. L. Lyra for critical comments and suggestions. This 
work was partially supported by FAPEAL (Funda\c{c}\~ao de Apoio \`a Pesquisa do 
Estado 
de Alagoas).


\section*{References}

\begin{thebibliography}{99}

\bibitem{kempe}J. Kempe, {\it Contemp. Phys.} {\bf 44}, 307 (2003).
\bibitem{Shenvi} N. Shenvi, J. Kempe, and K. Birgitta Whaley, {\it 
Phys. Rev. A} {\bf 67}, 052307 (2003).
\bibitem{Chakraborty}S. Chakraborty,L. Novo, A. Ambainis, Y. Omar, {\it 
Phys. Rev. Lett.} 
{\bf 116}, 100501 (2016) .
\bibitem{Portugal} Portugal R., {\it Quantum Walks and Search Algorithms} 
(New York: Springer) (2013).
\bibitem{Mohseni}M. Mohseni, P. Rebentrost, S. Lloyd, and A. Aspuru-Guzik,{\it 
The Journal of Chem. Phys.} {\bf 129}, 174106 (2008).
\bibitem{Schreiber} A. Schreiber, {\it et al}, {\it Phys. Rev. Lett.} {\bf 
104}, 050502 (2010).
\bibitem{Grafe} M. Gr\"afe, {\it et al}, {\it Sci. Rep.} {\bf 2}, 562 (2012).
\bibitem{Perets} Hagai B. Perets, {\it et al}, {\it Phys. Rev. Lett.} 
{\bf 100}, 170506 (2008).
\bibitem{Peruzzo} A. Peruzzo, {\it et al}, {\it Science} {\bf 329}, 1500 (2010).
\bibitem{Du}Jiangfeng Du, {\it et al}, {\it Phys. Rev. A} {\bf 67}, 042316 
(2003).
\bibitem{Manouchehri} K. Manouchehri and J. B. Wang, {\it J. of Phys. A: 
Math. and Theor.} {\bf 41}, (2008) 6.
\bibitem{Karski} M. Karski, {\it et al}, {\it Science} {\bf 325}, 174 (2009).
\bibitem{Preiss} P. M. Preiss, {\it et al}, {\it Science} {\bf 347}, 1229 
(2015).
\bibitem{Yin}Yue Yin, D. E. Katsanos, and S. N. Evangelou, {\it Phys. Rev. 
A} {\bf 77}, (2008) 022302.
\bibitem{Jackson}S. R. Jackson, T. J. Khoo, and F. W. Strauch, {\it Phys. Rev. 
A} {\bf 86}, 022335 (2012).
\bibitem{Wang}L. Wang, Li Wang, and Y. Zhang, {\it Phys. Rev. A} {\bf 90}, 
063618 (2014).
\bibitem{Peixoto} A.S. Peixoto, W.S. Dias, {\it Solid State Commun.} {\bf 
242}, 68 (2016).
\bibitem{loyd} Seth Lloyd, {\it J. Phys. Conf. Ser} {\bf 302}, 012037 (2011).
\bibitem{Childs} A. M. Childs, D. Gosset, Z. Webb, {\it 
Science} {\bf 339}, 791 (2013).
\bibitem{Berry} S. D. Berry and J. B. Wang, {\it Phys. Rev. A} {\bf 83}, 
042317 (2011).
\bibitem{Gamble} J. 
K. Gamble, M. Friesen,D. Zhou, R. Joynt, S. N. Coppersmith, {\it Phys. 
Rev. A} {\bf 81}, 
052313 (2010).
\bibitem{Dias}W. S. Dias, E. M. Nascimento, M. L. Lyra, and F.A.B.F. de Moura, 
{\it Phys. Rev. B} {\bf 76}, 155124 (2007).
\bibitem{Corrielli} G. Corrielli, et al., {\it Nature Communications} {\bf 
4}, 1555 (2013).
\bibitem{Thommen}Q. Thommen, J. C. Garreau, and V. Zehnle, {\it Phys. Rev. 
A} {\bf 65}, 053406 (2002).
\bibitem{Ivanov} 
V. V. 
Ivanov, A. Alberti, M. Schioppo, G. Ferrari, M. Artoni, M. 
L. Chiofalo, G. M. Tino, , {\it Phys. Rev. Lett.} {\bf 100}, 
043602 (2008).
\bibitem{Haller} 
E. Haller, R. Hart, M. J. Mark, J. G. Danzl, L. Reichsollner, H. 
C. Nagerl, {\it Phys. Rev. Lett.} {\bf 104}, 
200403 (2010).
\bibitem{Caetano} R. A. Caetano and M. L. Lyra, {\it Phys. Lett. A} {\bf 
375}, 2770 (2011).
\bibitem{DiasACDC} W. S. Dias, F. A. B. F. de Moura, and M. L. Lyra,
{\it Phys. Rev. A} {\bf 93}, 023623 (2016).
\bibitem{Zheng} Yi Zheng and Shi-Jie Yang, {\it New J. Phys.} {\bf 18}, 013005 
(2016).
\bibitem{Urano}Taeko I. Urano, Hiro-o Hamaguchi, {\it Chem. Phys. Lett.} 
{\bf 195}, 287 (1992).
\bibitem{Moore} 
F. L. Moore, J. C. Robinson, C. F. Bharucha, B. Sundaram, M. G. Raizen, {\it 
Phys. Rev. Lett.} {\bf 75}, 4598 
(1995).
\bibitem{Hu}Q. Hu, {\it et al}, {\it Phys. Rev. E} {\bf 71}, 031914 (2005).
\bibitem{Dong}G. Dong, W. Lu, and P. F. Barker, {\it Phys Rev. A} {\bf 69}, 
013409 (2004).
\bibitem{Beebe}S. J. Beebe and Karl H. Schoenbach, {\it Journal of 
Biomed. and Biot.} {\bf 2005}, 297 (2005).
\bibitem{Arlinghaus}S. Arlinghaus and M. Holthaus, {\it Phys. Rev. A} {\bf 
84}, 063617 (2011).
\bibitem{Silva} A. R. C. B. da Silva, F. A. B. F. de Moura, W. S. Dias,  
{\it Solid State Comm.} {\bf 236}, 12 (2016).
\bibitem{claro} J. F. Weisz and F. Claro, {\it J. Phys.: Condens. Matter} 
{\bf 15}, 3213 (2003).
\bibitem{Shepelyansky} D. L. Shepelyansky, {\it Physical Review Letters} 
{\bf 73}, 2607 (1994).
\bibitem{Dias2} W.S. Dias and M.L. Lyra, {\it Physica A} {\bf 411}, 35 (2014).
\bibitem{Islam} Rajibul Islam, {\it et al}, {\it Nature} {\bf 528}, 77 (2015).


\end{thebibliography}
\end{document}